\documentstyle[epsfig]{l-aa}

\def\ltsima{$\; \buildrel < \over \sim \;$}
\def\simlt{\lower.5ex\hbox{\ltsima}}
\def\gtsima{$\; \buildrel > \over \sim \;$}
\def\simgt{\lower.5ex\hbox{\gtsima}}

\def \rsun {\ifmmode$R$_{\odot}\else R$_{\odot}$\fi}

\def \hcm {\hbox {\ifmmode $ H atoms cm$^{-2}\else H atoms cm$^{-2}$\fi}}
\def\approxgt{\mathrel{\hbox{\rlap{\lower.55ex \hbox {$\sim$}}
        \kern-.3em \raise.4ex \hbox{$>$}}}}
\def\approxlt{\mathrel{\hbox{\rlap{\lower.55ex \hbox {$\sim$}}
        \kern-.3em \raise.4ex \hbox{$<$}}}}

\newcommand {\sax} {{\it BeppoSAX}}

\begin{document}
\thesaurus{ }

%
   \title{\sax~ observation of 3C273: broadband spectrum and detection
of a low-energy absorption feature}

   \author{P. Grandi\inst{1}\and M. Guainazzi\inst{2,1} \and
     T. Mineo\inst{3} \and A. N. Parmar\inst{4} \and
     F. Fiore\inst{2,5} \and A. Matteuzzi\inst{2} \and  F. Nicastro\inst{1} 
    \and G. C. Perola\inst{6}  \and L. Piro\inst{1} \and 
   M. Cappi\inst{7,8} \and G. Cusumano\inst{3} \and F. Frontera\inst{9} 
   S. Giarrusso\inst{3} \and E. Palazzi\inst{5}  \and S. Piraino\inst{3}}

\institute{{Istituto di Astrofisica Spaziale, C.N.R., Via Enrico Fermi, 21
I-00044 Frascati (RM), Italy}
\and
 {\sax~Science Data Center , c/o Nuova Telespazio, Via 
   Corcolle 19 I-00131 Roma, Italy }
\and
{Istituto di Fisica Cosmica ed Applicazioni dell'Informatica C.N.R., Via Ugo La Malfa 153, I-90146, Palermo, Italy}
\and 
{Astrophysics Division, Space Science Department of ESA, ESTEC, Postbus 299, 2200 AG Noordwijk, The Netherlands}
\and
{Osservatorio Astronomico di Roma, 
Via dell'Osservatorio, I-00044, Monteporzio Catone, Italy}
\and
{Dipartimento di Fisica ``E.Amaldi'', Universit\`a degli Studi ``Roma 3'', Via della Vasca Navale 84, I-00146, Roma, Italy}
\and
  {Istituto di Tecnologie e Studio delle radiazioni Extraterrestri C.N.R., Via P.Gobetti 101, I-40129, Bologna, Italy}
\and
{The Institute of Physical and Chemical Research (RIKEN), Wako-shi, Saitama, 350-01, Japan}
\and
{Dipartimento di Fisica, Universit\`a di Ferrara, Via Paradiso 11, I-44100 Ferrara, Italy}}
   \offprints{P. Grandi}

   \date{Received 1997 June 3; accepted 1997 June 26}

  \maketitle

 \markboth{P.Grandi et al.}{\sax~observation }

   \begin{abstract}

    We report the results of a 3C273 observation performed during the  
    Science Verification Phase (SVP) of the \sax~satellite.
The broad-band spectrum is well represented by a power-law 
 between $\sim$ 1~keV and 200~keV. The spectral slope is flat 
($\Gamma\sim 1.6$), with a weak emission line at $\sim 6.4$ keV  (rest frame) 
of EW $\sim 30 $ eV.      
    Below 1 keV, a deviation from a power-law
due to an absorption feature plus a soft component is present. 
This is the first time that 
a feature in absorption at $\sim 0.5$ keV (observer frame) 
is unambiguously detected in 3C273. 
          
    \keywords{X-ray: observations -- Quasar: 3C273}

   \end{abstract}

\section{Introduction}

3C273 is a bright, relatively nearby ($z=0.158$) quasar and therefore
has been very
well studied at all wavelengths from radio to $\gamma$-rays.
It is a core-jet radio source characterized by a compact nucleus and a jet
showing superluminal motion.
Multifrequency campaigns have recently shown that at least 3 maxima 
characterize the Spectral Energy Distribution (SED)
of this object, the emission peaking
in the IR, in the UV and in the $\gamma$-rays (Lichti et al. 1995,
von Montigny et al. 1997).
The complexity of the spectrum indicates that several physical mechanisms
contribute to the continuum.
It is generally believed that synchrotron emission dominates the spectrum 
from radio to IR frequencies (Robson et al. 1993),
thermal emission is responsible for the UV bump (Ulrich et al. 1988) 
and inverse Compton emission produces 
the observed spectrum from a few keV up to
GeV energies (von Montigny et al. 1997 and references therein). 
In the 2--10~keV region, 3C273 is represented by a power-law which 
extends up to $\sim 1$ MeV,  where a break has been recently detected by 
OSSE (McNaron-Brown et al. 1995). The presence of a reflection 
component,
similar to that observed in Seyfert galaxies, was 
suggested by GINGA data (Williams et al. 1992),
but has never been confirmed. On the contrary, the iron line at 6.4 keV
(rest frame) 
detected by GINGA, when the source was at a very low 
flux level, was later confirmed by ASCA (Cappi \& Matsuoka 1996).
At low energies, EXOSAT and ROSAT found evidence of a strong excess,
usually parameterized by power-law or thermal 
models (Turner et al. 1990; Leach et al. 1995; Laor et al. 1994).
Here we will show that the low-energy
spectrum of 3C273 is actually more complex
than implied by previous observations. 
Below 1 keV, the \sax~ data show a clear feature in absorption
and a soft component.
This is the first time that an absorption feature at $\sim 0.5$ keV
(observer frame) is unambiguously detected in the 3C273 spectrum.

\section{Observations and Data Reduction}

3C273 was observed during the Science Verification Phase (SVP) 
from 1996 July 18 to July 21 with the NFI
of the \sax~satellite.
The NFI consist of  four co-aligned detectors: 
the Low Energy Concentrator Spectrometer (LECS: 0.1-10 keV; Parmar 
et al. 1997), 
the Medium Energy Concentrator Spectrometer (MECS: 1-10 keV) 
(Boella et al. 1997), 
the High Pressure Gas Scintillation Proportional Counter (HPGSPC: 4-120 keV;
Manzo et al. 1997) and 
the Phoswich Detector System (PDS: 15-300 keV; Frontera et al. 1997).

All instruments 
operated in default configuration and data were telemetred in direct
modes. Standard data selection criteria were
applied\footnote{see {\verb!http://www.sdc.asi.it/software/cookbook!}
as a reference about data analysis software, reduction/analysis
procedures and calibrations}.
LECS, MECS and PDS data
were reduced  using {\sc Saxdas v.1.1.0} package.
For HPGSPC data, {\sc Xas v.2.0.1}
package (Chiappetti 1996) was used.
Data analysis was performed using the {\sc Xanadu} package.
Total usable exposure time was 
12 ksec, 131 ksec, 32 ksec and 64 ksec for
LECS, MECS, HPGSPC  and PDS, respectively.
The short LECS exposure is
due to the LECS being operated only during spacecraft night.
In the LECS image the source was about 2' away from the center of the field
of view and was partially obscured by a coarse strongback structure.

The LECS and the three MECS spectra were accumulated on 
circular regions of 8' and 4' radius, respectively.
Background spectra were extracted from blank fields observations
in the same position of  the source.
HPGSPC and PDS net source spectra and light curves were
produced by subtracting the off- from the on-source products.
A discussion of the systematics associated with such a procedure
can be found in Manzo et al. (1997) and Matt et al.
(1997) respectively.

The spectra of all the instruments
were rebinned to achieve at least 20 counts per bin, in order
to ensure the applicability of $\chi^2$ test in the spectral fits.
Publicly available matrices (1996, December 31 release)
were used for all the instruments but the LECS, for which an {\it ad hoc}
effective area was used to account for the  offset of the source
centroid. 

\section{Results}

A monotonic trend was observed 
in the MECS 
light curve (1.5-10 keV) with the intensity decreasing by $\sim 15\%$
during the pointing.
No  associated spectral variability was detected.

\subsection{The continuum shape above 0.8 keV}

We fitted a simple power-law model, absorbed by a neutral column density $N_H$,
separately to the spectra of the four detectors.
The results are summarized in Table~1.
The column density was constrained to the Galactic value 
($N_H^{Gal} = 1.68 \times 10^{20} \ cm^{-2}$, Savage et al. 1993). 
Hereafter statistical uncertainties
are 90\% confidence
level for one interesting parameter ($\Delta \chi^2 = 2.706$).

\begin{table}
\begin{center}
\begin{flushleft}
\caption[] {Power-law fits to single detectors spectra. $\Gamma$ is the
photon spectral index. Column
density has been held fixed to the Galactic value $N_H^{Gal} =
1.68 \times 10^{20} \ cm^{-2}$.  N/$N_{MECS}$ is the relative
normalization  at @ 1 keV from 
the simultaneous fit of all the instruments.}

\begin{footnotesize}
\begin{tabular}{lccc}
\noalign {\hrule}
\multicolumn{1}{c}{Detector}&
\multicolumn{1}{c}{$\Gamma$}&
\multicolumn{1}{c}{$\chi^2$/d.o.f}&
\multicolumn{1}{c}{$N/N_{MECS}$}\\
&&&\\
LECS (0.12--9.5~keV)	& $1.48\pm0.03$	& $329/265$	& ...	\\
LECS (0.8--4~keV)	 	& $1.52^{+0.07}_{-0.08}$ 	& 153/141
 	& $0.76\pm0.04$\\
MECS (1.5--10~keV)      & 1.57$\pm0.01$                 & 191/181
& 1.0 \\
HPGSPC (7--60~keV) 		& 1.5$\pm0.3$           	
& 239/211     	& $1.07\pm0.09$ \\ 
PDS (20--200~keV) 		& 1.60$^{+0.09}_{-0.07}$ 	
& 84/122    	& 0.73$\pm0.04$\\
\noalign {\hrule}
\end{tabular}
\end{footnotesize}
\end{flushleft}
\end{center}
\end{table}
This simple parameterization is a fairly good representation of
the spectral shape in the MECS, HPGSPC and PDS. 
Spectral indices are consistent with
each other within the statistical uncertainties.
In the LECS data, instead, 
a deviation from the simple power-law
behaviour
is present below $E \sim 0.8$ keV,
and is responsible for a rather high $\chi^2$;
spectral behaviour in this band will be discussed
in $\S3.2$.
The LECS spectral slope is flatter that the MECS one 
in the overlapping energy band
($\Gamma^{LECS}_{2-9.5 \ keV}=1.34\pm0.07$). 
This effect has been revealed in several sources observed by \sax~
so far and is probably due to a combination of penetration in the
driftless LECS gas cell (see Parmar et al. 1997) and obscuration
by a window support structure rib.
If the LECS data are fitted in the 
0.8--4~keV energy range, the spectral index is consistent the MECS slope,
within the statistical uncertainties (see Table~1).

We then fitted  the  power-law model simultaneously 
to the data of the four detectors (0.8--200~keV),
restricting the LECS data to the 0.8-4 keV band. 
The resulting normalization factors are different 
in the four instruments: their values with respect to the MECS 
(N/M$_{MECS}$ in Table~1)
are consistent with those obtained in the calibration performed
on the Crab nebula (Cusumano et al. 1997).
The fit is acceptable, $\chi^2/d.o.f= 628/621$, (see Figure 1), the spectral 
photon index is $\Gamma=1.57\pm0.01$ and the unabsorbed
flux in 2--10~keV is $F=(7.1\pm0.2) \times 10^{-11} \ erg \ cm^{-2} \ s^{-1}$.
\begin{figure}
\label{fig1}
\epsfig{figure=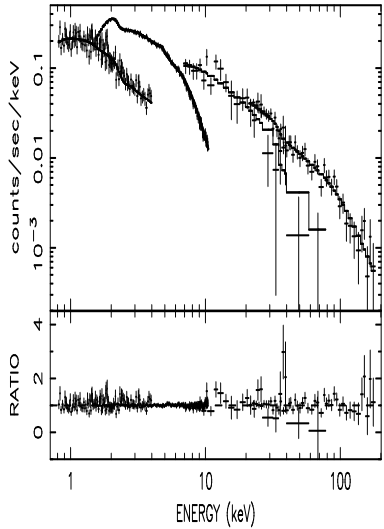,height=8.5cm,width=5.5cm,angle=-90}
\caption{Broadband 3C273 spectrum {\it (upper panel)} and the 
residuals 
{\it (lower panel)} when a simple absorbed ($N_H = N_H^{Gal}$) 
power-law model is applied in the 0.8--200~keV energy range}.
\end{figure}

\subsection{Spectral features below 0.8 keV}

As mentioned in the previous section, strong deviations from a  simple power 
law model are present in the 3C273 spectrum below 0.8 keV.
When a  power law is fitted to the LECS data (0.12-4.0 keV), 
fixing the photon index to the average broadband slope 
($\Gamma=1.57$) and the cold absorber to the Galactic value, 
the $\chi^2$ is unacceptable ($\chi^2$/d.o.f = 259/196).
The inspection of the residuals clearly shows
a photon deficit around E$\sim 0.6$ keV and an excess  emission below
$\sim$ 0.3 keV (Fig.2, left panel).

\begin{figure}
\label{fig2}
\epsfig{figure=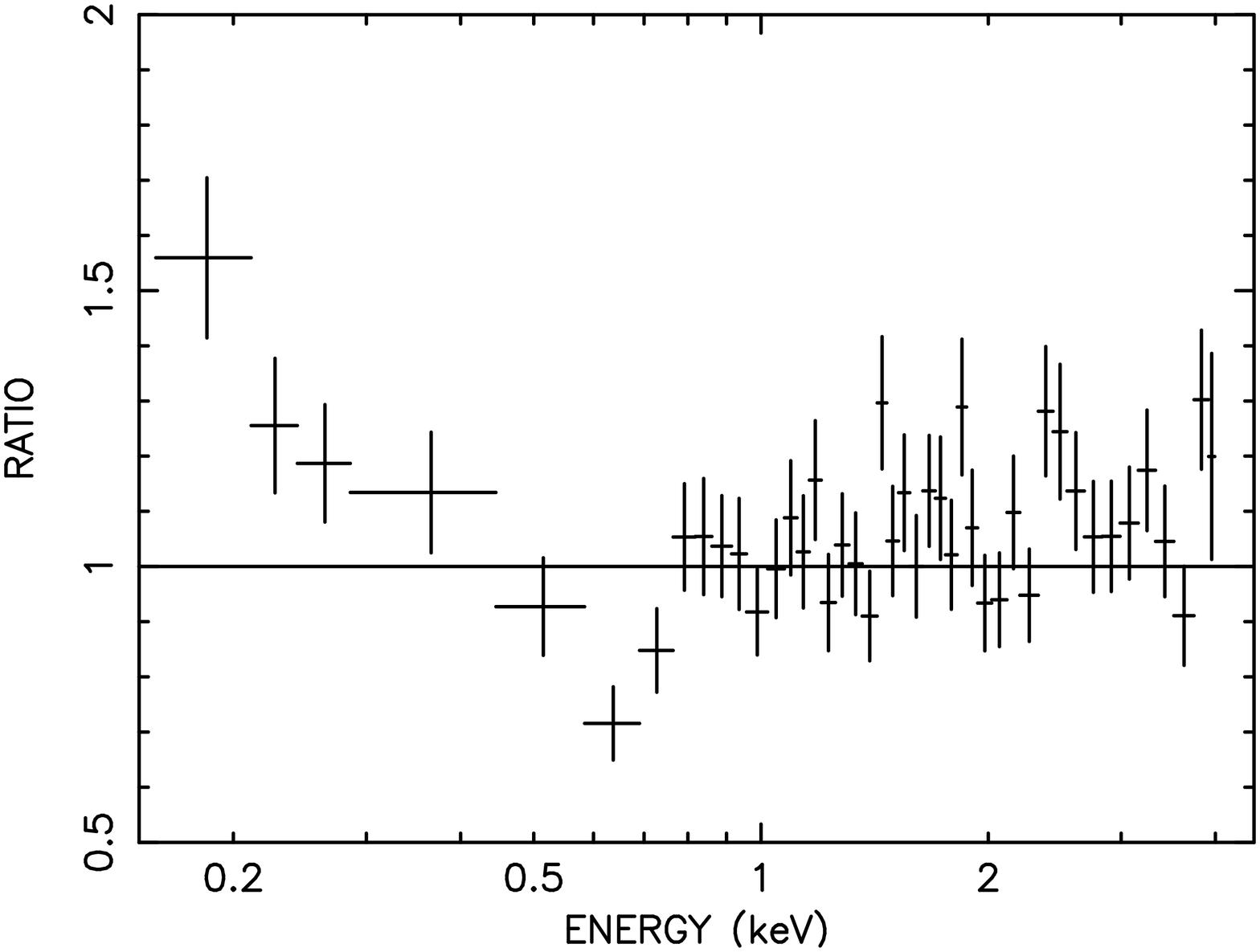,height=4.0cm,width=5.5cm, angle=-90}
\epsfig{figure=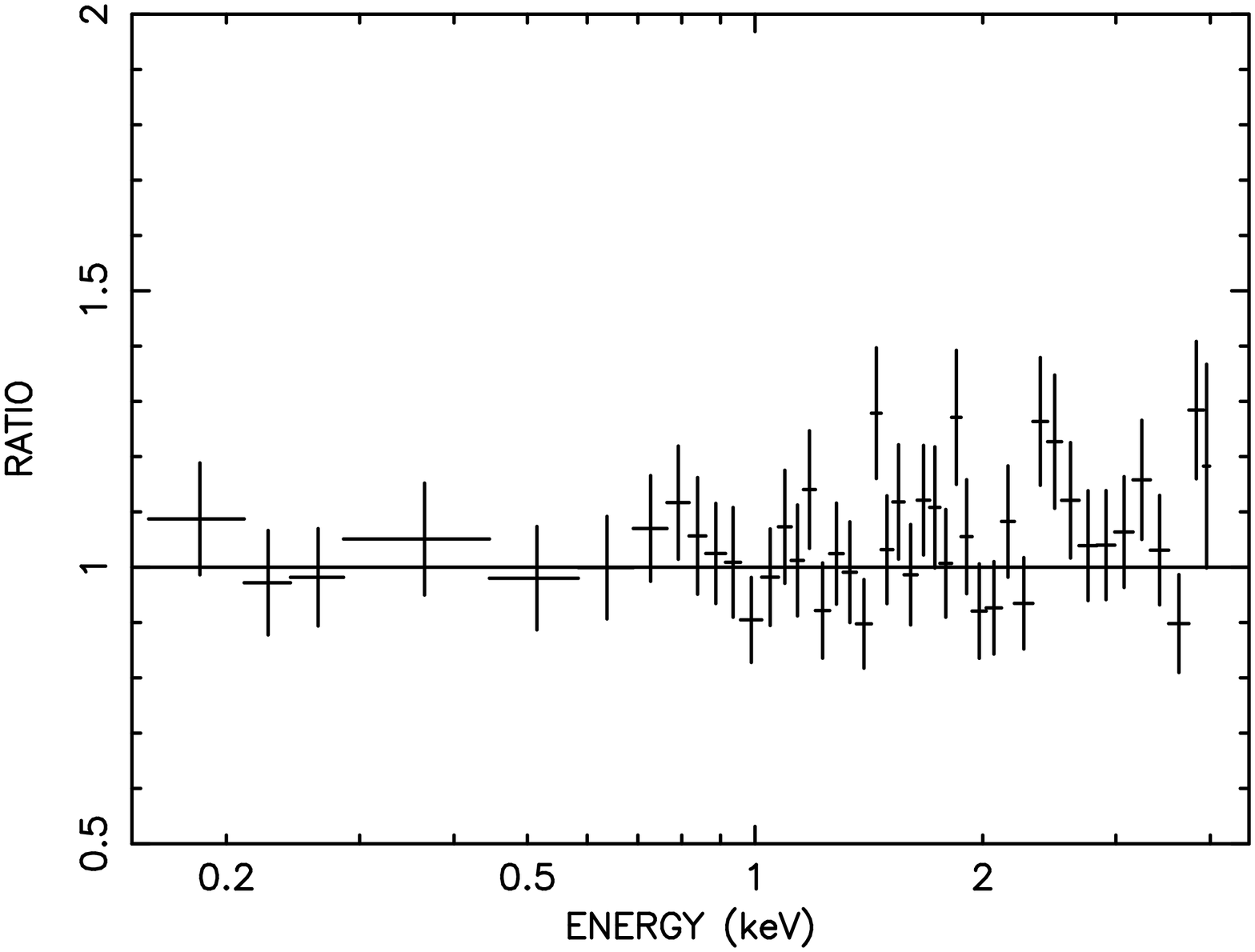,height=4.0cm,width=5.5cm, angle=-90}
\caption{An absorption feature and a soft excess are clearly evident 
in the LECS  residuals when a  power  law ($\Gamma=1.57$, 
N$_H$=N$^{Gal}_{H}$) is fitted the the data ({\it left panel}). 
Data are well parameterized by a broken 
power-law plus  a notch ({\it right panel}) }
\end{figure}
We first parameterized the continuum with a broken power-law
($\Gamma_{hard}=1.57$) and then added an absorption feature using
an absorption edge or a rectangular trough 
profile (model {\sc notch} in {\sc Xspec}). We fixed the notch covering
fraction equal to 1 since it was not possible to simultaneously determine
meaningful constraints on its depth and width. 
The broken power law alone can 
not adequately reproduce the LECS data (see Table 2). Adding 
either an edge or a notch yields a statistically significant 
improvement $\chi^2 = 28$ (Fig.2, right panel) in both cases. 
Equally acceptable fits are obtained 
when a power law ($\Gamma_{soft} \simeq 5.0 -  5.5$) or a blackbody
($T \simeq 20$ eV) are used to describe the soft excess. Similar absorption 
best-fit parameters are obtained. In Table~2
the best-fit parameters are quoted for the broken power-law case
(hereafter energies are quoted in the source rest frame).

\begin{table}
\begin{center}
\begin{flushleft}
\begin{footnotesize}
\caption[] {Best fit parameters of the soft emission when a broken power-law 
model + a feature in absorption are applied  to the LECS data in the 
0.1--4~keV band. 
 $E_{break}$
was fixed to the best-fit value 0.30 keV to calculate 
the statistical uncertainties. $N_H = N_H^{Gal}$, 
$\Gamma_{hard}=1.57$ (fixed). }
\begin{tabular}{lcccc}
\noalign {\hrule}
\multicolumn{1}{l}{}&
\multicolumn{1}{c}{$\Gamma_{soft}$}&
\multicolumn{1}{c}{E$^a$}&
\multicolumn{1}{c}{$\tau$/Width}&
\multicolumn{1}{c}{$\chi^2$/d.o.f}\\
&& (keV)& (/eV) & \\
&&&&\\
BPL   & $3.0^{+0.4}_{-0.5}$ &...& ...& 238/194\\
BPL + Edge & $2.7^{+0.4}_{-0.5}$& $0.61^{+0.09}_{-0.05}$& $0.7\pm0.3$&210/192\\
BPL + Notch & $2.8^{+0.4}_{-0.5}$& 0.81$^{+0.04}_{-0.06}$& 64$\pm22$&210/192\\
\noalign {\hrule}
\multicolumn{5}{l}{$^a$ - Rest frame energy}\\
\end{tabular}
\end{footnotesize}
\end{flushleft}
\end{center}
\end{table}

\subsection{Iron Line Emission}

In the MECS spectrum there is evidence
of an emission line around $E \sim 5.4 \ keV$.
If a Gaussian profile is added
to the power-law model in the MECS spectrum,
the $\chi^2$ is reduced by $\Delta \chi^2 = 15$,
with best-fit (rest frame) parameters: $E = 6.22 \pm 0.12$ keV and $EW =
30 \pm 12$ eV.
In order to test the influence of minor
miscalibration around the nearby Xenon edge ($E_{Xe} \simeq 4.7 \ keV$),
we have divided the MECS 3C273 spectrum by the Crab spectrum and
inspected the residuals around the line energy.
The line structure is still present and a Gaussian
fit  yields an $EW = 26 \pm 13 \ eV$.
The line can be naturally explained
as K$_{\alpha}$ fluorescence from neutral or mildly ionized
iron, in agreement with the outcome by GINGA
(Williams et al. 1992) and ASCA (Cappi $\&$ Matsuoka 1996).

\section{Discussion}
We have presented the results of the {\it Beppo}SAX observation of 3C273 
during the SVP. 
The overall continuum is well represented by a simple power-law from 
$\sim 1$ keV to 200 keV. A weak iron line is observed 
in the MECS spectrum.
Below 1 keV, The LECS data show a soft excess and a feature in 
absorption. 

A non-thermal origin for the hard X-ray/$\gamma$-photons has been 
suggested so far. 
Arguments based on the observed 
$\gamma$-rays luminosity and rapid X-ray variability 
show that the radiation must be relativistically beamed
(Lichti et al. 1995).
Relativistic electrons associated with 
the jet upscatter
lower energies photons located in the same jet (synchrotron radiation,
Maraschi et al. 1992; Bloom $\&$ Marscher 1993) or/and
coming from  the accretion disk (Dermer et al. 1992) 
or from the Broad Line Region (Sikora et al. 1994).
If this picture is a realistic representation of 3C273, 
the weak line (EW$\sim 30 $ eV) can be interpreted 
as a typical $K_{\alpha}$ iron line from accretion disk,   
diluted by a strong Doppler-enhanced continuum.
An underlying 
'Seyfert like' component is then expected to 
take part  to the X-ray emission.
The amount of  such a contribution  
can be deduced by comparing 
the EW of the observed line with the  typical EW ($\sim 200$ eV) 
of the same line in Seyfert 1 galaxies (Nandra and Pounds 1994).
It turns out to be  $\sim 15\%$ at 6 keV.

We explored whether the Seyfert-like component  can at least
partially account for the observed soft excess. 
The LECS+MECS  spectra were  modeled  with a double power-law plus a notch
with fixed Galactic absorber.
One power law slope was left to vary, the other one 
was fixed  to the typical Seyfert 1 photon index ($\Gamma_{unbeamed}=2$).
The fit result is: 
$\Gamma_{beamed}=1.52^{+0.05}_{-0.04}$, $N_{unbeamed}/N_{beamed}=10\pm6\%$ @
6.4 keV $(\chi^2$=411/381 d.o.f.).
The notch parameters are consistent with the ones in Table~2.
We conclude that an unbeamed  underlying component could account for the
observed soft excess, although a residual contribution from the hard tail 
of the UV bump cannot be completely ruled out (Fig. 3).

\begin{figure}
\label{fig3}
\epsfig{figure=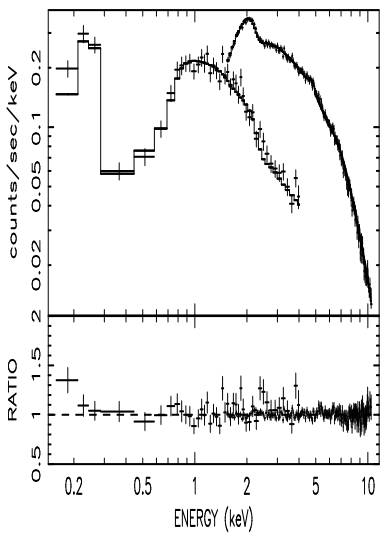,height=8.5cm,width=5.5cm,angle=-90}
\caption{Spectrum ({\it upper panel}) and residuals ({\it lower panel}) 
when a double power-law model is applied to
the  spectrum of 3C273 in the 0.12--10~ keV band.}
\end{figure}

We also report the detection of an absorption feature in the
soft X-ray spectrum of 3C273. Such a feature can be modelled either with
an absorption edge or with a notch. 
No discrimination between the two models can be made on statistical basis.

In the former case, the best-fit energy
is consistent with the OIII-OV ionization stages at the source rest frame.
We used Cloudy (version 90.01) to simulate absorption from gas 
characterized by a low ionization state, in order to reproduce 
the observed features.  
The ionizing continuum was described by a broken 
power law, as indicated by the \sax~data, with
the softer part (spectral index $\Gamma= 3$)
extrapolated to lower energies to represent the UV continuum.
The model can reproduce the observed feature, but, in addition,
predicts strong absorption due to
C (IV,V)  and N (IV,VI) and HII below 0.5 keV, in completely
disagreement with the excess emission present in the LECS data.

A possibility is that the absorption edge is produced by highly ionized
oxygen (OVII, OVIII), as usually observed in Seyfert 1 galaxies.
In this case, the gas is required to inflow ($v_{inflow}/c \sim 0.19-0.4$)
to explain the {\it redshifted} position of the feature 

Alternatively, the notch structure
might be due to highly ionized Oxygen (the most likely candidate being
the {\sc Oviii} Ly-${\alpha}$)  as suggested for PKS2155-304
(Canizares \& Kruper 1984; Madjeski et al. 1991, Giommi et al. 1997).
In such a case, high-velocity material 
($v_{outflow}/c\sim 0.15-0.30$) within the jet 
might produce the {\it blueshifted} absorption through.

\begin{acknowledgements}
We would like to acknowledge all the members of the \sax~team, whose work 
has allowed a successful launch and management of the mission.
We would like to also thank L. Maraschi for many  useful 
discussions.
\end{acknowledgements}

\end{document}